\documentclass[preprint,aps,showpacs,twocolumn]{article}
\usepackage{amssymb}
\usepackage{amsthm,amsmath}
\usepackage{graphicx,subfigure,xcolor}
\usepackage{lscape}
\usepackage{float}
\usepackage[margin=0pt,font=small,labelfont={bf,normalsize},labelsep=space]{caption}
\usepackage[numbers,sort&compress]{natbib}
\usepackage[colorlinks,citecolor=red,urlcolor=blue,linkcolor=red]{hyperref}
\makeatletter
\renewcommand{\@seccntformat}[1]{\csname the#1\endcsname.\quad}
\makeatother

\newcommand{\customfootnotetext}[2]{{
		\renewcommand{\thefootnote}{#1}
		\footnotetext[0]{#2}}}
	
\begin{document}
	\twocolumn[
	 \begin{@twocolumnfalse}
\title{\Large{\bf {Kinematic corrections and reconstruction methods for neutral Higgs boson decay to $b\bar{b}$ in 2HDM type-I at future lepton colliders }}}

\author{Majid Hashemi\textsuperscript{$\star$} and Elnaz Ebrahimi\textsuperscript{$\dagger$}\\
\emph{Physics Department, College of Sciences, Shiraz University, Shiraz, 71946-84795, Iran}}
\maketitle
\begin{abstract}
	\begin{center}
	\begin{verse}
	In this paper, an approach for neutral Higgs boson searches is described based on 2HDM type-I at electron-positron linear colliders operating at $ \sqrt{s}=1$ TeV. The beam is assumed to be unpolarized and fast detector simulation is included. The signal process produces a fully hadronic final state through $ e^{+}e^{-}\rightarrow AH\rightarrow b\bar{b}b\bar{b} $ where both CP-even and CP-odd Higgs bosons ($H$ and $A$) are assumed to decay to a pair of $b$-jets. Several benchmark scenarios are introduced as the baseline for the analysis taking $ m_{H/A}$ in the range 150--300 GeV. In order to avoid Higgs boson conversion $A \to ZH$, Higgs boson masses are chosen with $m_A-m_H < m_Z$. It is shown that with a proper kinematic correction applied on final state $b$-jet four momenta, true combinations of $b$-jets can be found for simultaneous reconstruction of both Higgs bosons through $b\bar{b}$ invariant mass calculation. Results show that observable signals can be achieved with statistical significance exceeding $5\sigma$ well before the target integrated luminosity of 8 $ab^{-1}$.\\
	\end{verse}	
	\end{center}
\end{abstract}
 \end{@twocolumnfalse}]
\customfootnotetext{$\star$}{email:majid.hashemi@cern.ch}
\customfootnotetext{$\dagger$}{ebrahimielnaz69@gmail.com}
\section*{\large{Introduction}}
\paragraph{}The Standard Model of particle physics (SM) is one of the most precisely tested theories which has been verified in various experiments. After the discovery of the electroweak gauge bosons, there has been extensive search for the missing key element of the standard model, i.e., the Higgs boson, $\mathit{h_{SM}}$ \cite{Spontaneous symmetry Higgs,kibble,Broken symmetries Higgs,symmetries Higgs,Englert,Guralink}. The result of these searches is the observation of a new boson at the Large Hadron Collider (LHC) by the two collaborations ATLAS and CMS \cite{ATLAS,CMS}. 

The properties of the observed boson show reasonable compatibility with SM predictions as verified at LHC \cite{LHC1,LHC2,LHC3,LHC4,LHC5,LHC6,LHC7,LHC8}. However, these measurements still allow possible ext\texttt{}ensions to the Higgs sector such as the two Higgs doublet model \cite{Branco CP,Lee,Glashow} which is the basis for several beyond SM scenarios such as supersymmetry \cite{supersymmetry,supersymmetric1,supersymmetric2}. 

The Two Higgs Doublet model (2HDM) has attracted attention even as a standalone model without necessarily embedding it in a supersymmetric theory. Since there are two Higgs boson doublets with complex fields, a total number of five Higgs bosons are predicted including the lightest Higgs boson $\mathit{h}$ (the SM-like Higgs boson), the heavy neutral CP-even(CP-odd) Higgs bosons $\mathit{H}$($\mathit{A}$) and the two charged Higgs bosons $\mathit{H^{\pm}}$.

After the discovery of the new boson, there has been extensive search for the extra Higgs bosons at LHC. The ATLAS collaboration has reported an analysis of $pp$ $\rightarrow A+X\rightarrow ZH+X $ followed by $Z\rightarrow \ell\ell$ and $H\rightarrow b\bar{b}$ at $\sqrt{s}$ = 13 TeV. The pseudoscalar Higgs decay is kinematically allowed if $m_A-m_H \geq m_Z$. There are two subsequent reports for this analysis based on integrated luminosity of 139 $fb^{-1}$ \cite{ATLAS06} and 36.1 $fb^{-1}$ \cite{ATLAS021}. Four types of 2HDM based on Higgs-fermion couplings have been analyzed and exclusion contours have been presented in parameter space of $m_A$ vs $m_H$ for various model parameter values. As mentioned before, the main characteristics of such analyses is the limitation in the Higgs boson mass parameter space which leaves the region of $m_A-m_H<m_Z$ untouched. 

The CMS collaboration has also reported similar analysis of $pp$ $\rightarrow A(H)\rightarrow ZH(A) $ with $Z\rightarrow \ell\ell$ and $H(A)\rightarrow b\bar{b}$ using integrated luminosity of 35.9$fb^{-1}$ \cite{CMS01}. There has been also analysis of heavy pseudoscalar $A$ boson decaying to $Z$ and SM-like $h$ boson with $m_h = 125$ GeV followed by $h \to b\bar{b}$ and $Z \to \ell^{+}\ell^{-}$ \cite{CMS03}. 

The overall conclusion from these searches is that the upper left region of the parameter space in $m_A$ vs $m_H$ plane defined by $m_A-m_H>m_Z$ is excluded up to $m_H\simeq 300$ GeV. The area of the excluded region, however, depends on the type of 2HDM. 

Before going to the analysis details, the theoretical framework of the analysis is described and the working points in the parameter space are introduced. These points represent example analysis scenarios which target the unexplored region of the parameter space by LHC. 
\section{Two Higgs Doublet Model}
The two \emph{SU(2)} doublets introduced in 2HDM contain complex fields resulting in eight degrees of freedom, three of which are eaten by the electroweak gauge bosons to receive their masses. Therefore five degrees of freedom remain leading to five physical Higgs bosons which are denoted as $h,~H,~A$ and $H^{\pm}$. The neutral Higgs masses are assumed to be in the same order as listed above, i.e., $h$ is the lightest Higgs boson which is considered to be SM-like and the other two neutral Higgs bosons are heavier.  

Each doublet has its own vacuum expectation value or "vev" ($v_1$ and $v_2$). They are related to the SM-like $v$ = 246 GeV through $v_1=v\cos\beta$ and $v_2=v\sin\beta$ resulting in the ratio $\tan\beta=v_2/v_1$ which is the free parameter of the model. 

In addition to $\beta$ parameter, there is also rotation angle, denoted by $\alpha$, which diagonalizes the mass-squared matrix of the neutral Higgs bosons. The two free parameters $\alpha$ and $\beta$ determine the Higgs-fermion couplings in various 2HDM types. 

The neutral scalar Higgs couplings with gauge bosons also depend on these parameters through $\sin(\beta-\alpha)$ ($h$-gauge) or $\cos(\beta-\alpha)$ ($H$-gauge) and therefore the lightest Higgs boson of the model acquires the same couplings with gauge bosons as those of $h_{SM}$ if $\sin(\beta-\alpha)=1$ which is the alignment limit \cite{alignment}. With this requirement, $H/A$-fermion couplings will solely depend on $\tan\beta$ or $\cot\beta$ and $h$-fermion couplings coincide their SM values which are $m_f/v$ with $m_f$ being the fermion mass \cite{Branco,Haber,Mrazek,Mahmoudi}. It should be noted that the alignment limit is naturally realized in the decoupling regime where the other Higgs bosons are decoupled by assuming that they are much heavier than the electroweak scale $v$ \cite{decoupling}. However, the Higgs boson masses under study in this work, are of $\mathcal{O}{(v}{)}$. Therefore the chosen scenario is the alignment limit without decoupling. 

The Yukawa Lagrangian for Higgs-fermion interactions can be written in this form:
\begin{eqnarray}\label{yukawa}
\mathcal{L}&=&v^{-1}(m_{d}d\bar{d}+m_{u}u\bar{u}+m_{\ell}\ell\bar{\ell})h\nonumber\\ &+&v^{-1}(\rho^{d}m_{d}d\bar{d}+\rho^{u}m_{u}u\bar{u}+\rho^{\ell}m_{\ell}\ell\bar{\ell})H\nonumber\\ &+&iv^{-1}(-\rho^{d}m_{d}\bar{d}\gamma_{5}d+\rho^{u}m_{u}\bar{u}\gamma_{5}u-\rho^{\ell}m_{\ell}\bar{\ell}\gamma_{5}\ell)A\nonumber \\
\end{eqnarray}

According to Tab. \ref{different type}, the heavy Higgs couplings acquire additional type dependent factor $\rho^{f}$ which can be used to distinguish the model type as well as the Higgs boson decay properties \cite{Aoki,Barger}.
\begin{table}[h]
	\centering
	\begin{tabular}{|ccccc|} \hline
		\multicolumn{5}{|c|}{2HDM Types}\\

		& I & II & III & IV \\ \hline
	
		$ {\rho^{d}} $ & ${\cot\beta}$ & ${-\tan\beta}$ & ${-\tan\beta}$ & ${\cot\beta}$ \\
		\hline
		$ {\rho^{u}} $ & ${\cot\beta}$ & ${\cot\beta}$ & ${\cot\beta}$ & ${\cot\beta}$ \\
		\hline
		$ {\rho^{\ell}} $ & ${\cot\beta}$ & ${-\tan\beta}$ & ${\cot\beta}$ & ${-\tan\beta}$ \\ \hline
	\end{tabular}
     \caption{The Higgs-fermion couplings in different 2HDM types.}
	\label{different type}
\end{table}

In Tab. \ref{different type}, $u(d)$ and $\ell$ denote the up(down)-type quarks and leptons. Type-III is also called ``\emph{Flipped}" and Type-IV is called ``\emph{lepton-specific}".
 
We have recently performed various studies of different 2HDM types at future lepton colliders. The main focus has been on Higgs boson pair production, through $e^+e^- \to HA$. 

In type-I, $H \to b\bar{b}$ has been shown to be the most promising decay channel with $A \to b\bar{b}$ \cite{HtypeI_1} or $A \to ZH$ with possible leptonic or hadronic decay of the $Z$ boson \cite{HtypeI_2,HtypeI_3,HtypeI_4}. The four $b$-jet final state through $H/A \to b\bar{b}$ has also shown discovery potential in the flipped type or type-III \cite{HtypeIII_1}.  In the lepton-specific type or type-IV, the leptonic decay channels, i.e., $H/A \to \tau\tau$ or $\mu\mu$, result in observable signals in parts of the parameter space as reported in \cite{HtypeIV_1,HtypeIV_2,HtypeIV_3}. 

In this work, the same Higgs boson pair production in the four $b$-jet final state is considered as the signal. If $m_{H/A}<2m_t$, the Higgs boson decay to top quark pair is kinematically forbidden. Given all Higgs-fermion couplings proportional to $\cot\beta$, the branching ratio of Higgs decay to fermions becomes independent of $\cot\beta$ as the common factors from partial decay rates and the total width cancel out. The remaining key parameter is thus the fermion mass resulting in dominant Higgs decay to $b$-jet pair.

It will be shown that using kinematic correction applied on final state four-momenta, a dramatic improvement of the results is obtained compared to those reported in \cite{HtypeI_1}. Moreover two possible approaches for the simultaneous reconstruction of the Higgs bosons are introduced and compared.  

 \section{Signal and background processes}
 
\paragraph{}The signal process is assumed to be the Higgs boson pair production producing four $b$-jet final state in the framework of type-I 2HDM, i.e.,  $e^{+}e^{-}\rightarrow AH\rightarrow b\bar{b}b\bar{b}$. Since Higgs-fermion couplings are proportional to the fermion mass and the common $\cot\beta$ factor cancels out in branching ratio calculations at tree level, both Higgs bosons predominantly decay to $b\bar{b}$ which is the heaviest accessible fermion pair, provided that the Higgs boson mass is below the top quark pair production threshold. 

The linear collider is assumed to be $e^+e^-$ collider operating at center-of-mass energy of $\sqrt{s}=1$ TeV which is realized at the upgrade phase of ILC with target integrated luminosity of 8 $ab^{-1}$ \cite{8ab,8ab2}. 

\begin{figure}[h]
	\centering
\includegraphics[width=0.49\textwidth]{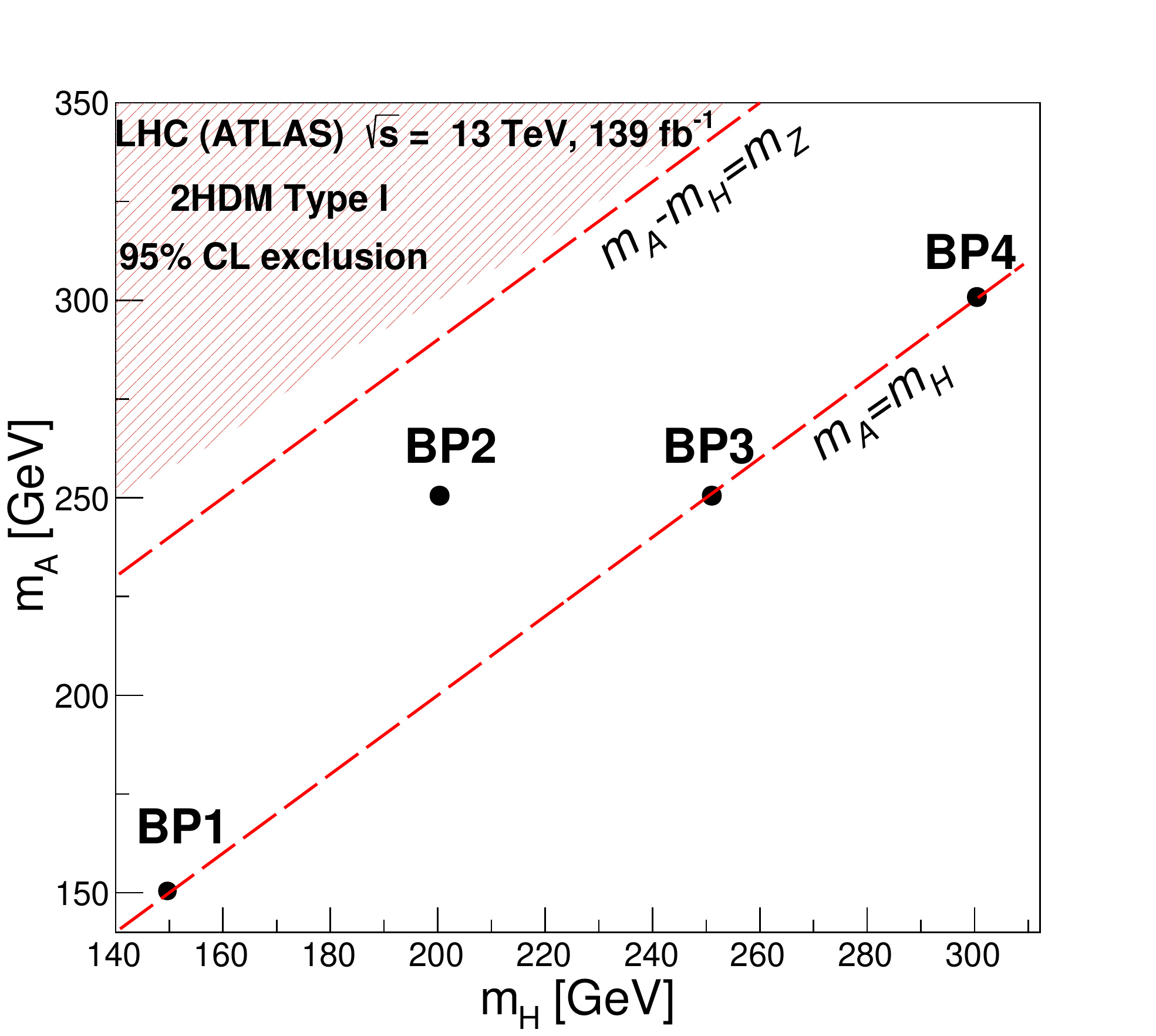}  
	\caption{The selected benchmark points in the parameter space of $m_A$ vs $m_H$. The current excluded region of LHC has also been shown based on \cite{ATLAS06}. \label{BPFigure}}
\end{figure}

For illustrative purposes, several benchmark scenarios are introduced and the analysis focuses on the selected points in the Higgs boson mass parameter space as shown in Fig. \ref{BPFigure}. The masses of the CP-even ($H$) and CP-odd ($A$) Higgs bosons are chosen to be in the region between the two dashed lines shown in Fig. \ref{BPFigure}. 

The analysis strategy is different from what is adopted by LHC experiments (ATLAS and CMS). They take $pp \to A+X$ as the signal followed by $A \to ZH$ decay which requires $m_A-m_H \geq m_Z$ \cite{ATLAS06}. This requirement limits their explorable region in $m_A$ vs $m_H$ plane which is shown with the upper dashed line in Fig. \ref{BPFigure}. In the current analysis, the Higgs boson fermionic decay, i.e., $H/A \to b\bar{b}$ is adopted for analysis. Therefore it is possible to explore regions in parameter space which are inaccessible by the current LHC analyses, i.e., those with $m_A - m_H < m_Z$. These points are well outside the excluded region of type-I 2HDM reported by LHC \cite{ATLAS06}. 

Table \ref{bps} shows parameter values for the four benchmark points BP1--BP4. The $\tan\beta$ parameter is set to 10 and $\sin(\beta-\alpha)$ is required to be 1 for all scenarios. The Higgs potential mass parameter $m^2_{12}$ is determined by searching for values which satisfy the theoretical requirements of potential stability \cite{Deshpande}, unitarity \cite{Unitarity,Unitarity2,Unitarity3} and perturbativity. Therefore, for each benchmark point, a range of allowed $m^2_{12}$ values is obtained as shown in Tab. \ref{bps}.

All calculations related to the parameter values and theoretical requirements as well as the Higgs boson branching ratio of decays are performed with the help of \texttt{2HDMC 1.8.0} \cite{2HDMC,2HDMC2,2HDMC3}. Agreement with experimental results is confirmed by embedding \texttt{HiggsBounds 5.9.0} \cite{hb1,hb2,hb3,hb4,hb5} and \texttt{HiggsSignal 2.6.0} \cite{hs1,hs2,hs3} in \texttt{2HDMC 1.8.0} where the selected benchmark points are checked to be not in the excluded regions reported by LHC and TeVatron experiments. 

\begin{table*}[h]
 	\centering
 	\begin{tabular}{|c|cccc|}\hline
 		
	Higgs bosons mass	& BP1 & BP2 & BP3 & BP4 \\
		\hline
 		$ {m_{h}} $  &\multicolumn{4}{c|}{125}  \\
 		\hline
		$ {m_{H}} $ & ${150}$ & ${200}$ & ${250}$ & ${300}$ \\
 	\hline
 		$ {m_{A}} $ & ${150}$ & ${250}$ & ${250}$ & ${300}$ \\
 		\hline
 		$ {m_{H^{\pm}}} $ & ${150}$ & ${250}$ & ${250}$ & ${300}$ \\
 		\hline
 		$ {m_{12}^{2}} $ & ${1987-2243}$ & ${3720-3975}$ & ${5948-6203}$ & ${8671-8926}$ \\
 		\hline
 		$ {\tan\beta} $ &\multicolumn{4}{c|}{10}\\
 		\hline
 		$ {\sin(\beta-\alpha)} $  &\multicolumn{4}{c|}{1} \\\hline
 	\end{tabular}
	\caption{The selected benchmark points of the analysis and parameter values.}
 	\label{bps}
 \end{table*}
 
Since, the signal final state consists of four $b$-jets, any SM process with the same final state should be regarded as the background. Moreover, detector effects, $b$-tagging fake rate and the final state radiation can also be a source of background. Therefore, Drell-Yan $ Z/\gamma^* $ (single neutral gauge boson), $ZZ$ (gauge boson pair) and $t\bar{t}$ (top quark pair) are the main background production processes.

The so called overlay hadronic background from photon interactions, i.e., $\gamma \gamma \to$ hadrons is not simulated in the analysis. However we follow the same approach as adopted by the CLIC collaboration reported in \cite{overlay} by adding the jet momentum smearing to account for the effect of the hadronic overlay on the jet reconstruction. The jet smearing at 1.5 TeV collisions proposed in \cite{overlay} assumes $1\%$ and $5\%$ relative smearing applied to the jet momentum with $|\eta|<0.76$ and $|\eta| \ge 0.76$ respectively. We perform a rough tuning of the above values to $0.7\%$ and $3\%$ in the corresponding pseudorapidity bins for 1 TeV collisions. 

\section{Analysis strategy for Higgs boson reconstruction}
In this section, the software setup for event generation, cross section calculation, detector response simulation and analysis is presented in detail including package versions which are all the current latest versions. 

The $b$-tagging algorithm and kinematic corrections applied on $b$-jets based on full four momentum conservation are described in the next sub-sections. The analysis details are then presented where two approaches for finding the true combinations of final state objects are described and the best approach is adopted. 
\subsection{Analysis software setup}
The signal and background generation is performed with the use of \texttt{WHIZARD 3.0.0-$\beta$} \cite{whizard1,whizard2} including the beam spectrum and initial state radiation (ISR). The beam spectrum file is taken from the official package repository \cite{ilc1000}. The generated event files are stored in \texttt{LHEF} format \cite{lhef} and passed to \texttt{PYTHIA 8.3.03} \cite{pythia} for the multi-particle interaction, showering and final state radiation (FSR). The \texttt{PYTHIA 8.3.03} output is used by \texttt{DELPHES 3.4.2} \cite{DELPHES} for detector response simulation using \texttt{ILCgen} detector card proposed for ILC. 

For the detector coordinate system we use the azimuthal angle $\phi$ and pseudorapidity defined as $\eta=-\ln{\tan({\theta/2})}$ where $\theta$ is the polar angle with respect to the beam axis.

The detector acceptance implemented in \texttt{ILCgen} card includes full pseudorapidity coverage for charged tracks with $p_T,\eta$ dependent momentum smearing and different tracking efficiencies in bins of $|\eta|<1.83$, $1.83<|\eta|<2.65$, $2.65<|\eta|<3$ and $|\eta|>3$ excluding track $p_T<0.1$ GeV. 

The ECAL and HCAL resolutions are assumed to be $\sigma/E=\sqrt{{(0.01)}^2+{(0.17)}^2/E}$ and $\sigma/E=\sqrt{{(0.017)}^2+{(0.45)}^2/E}$. The acceptances of the two calorimeter subdetectors are assumed to be $|\eta|<4$ and $|\eta|<3.8$ for ECAL and HCAL respectively.  

The jet reconstruction is performed using \texttt{FASTJET 3.3.4} \cite{FASTJET1,FASTJET2} (embedded in \texttt{DELPHES}) with anti-$k_{t}$ algorithm \cite{anti-$K_t$} with the jet cone size $\Delta{R}=\sqrt{(\Delta{\eta})^{2}+(\Delta{\phi})^{2}}=0.5$.

The \texttt{ILCgen} card contains detailed $b$-tagging scenarios with the $b$-tagging efficiencies implemented as a function of the jet energy and pseudorapidity with average $b$-jet identification efficiencies of 80$\%$, 70$\%$ and 50$\%$. The fake rate has also been included for all $b$-tagging scenarios.

The detector simulation output is stored in \texttt{ROOT} files which are created using \texttt{ROOT 6.22/08} \cite{ROOT} and serve as the datasets for the final numerical and graphical analysis.

\subsection{Cross sections}
The signal and background cross sections are obtained using both \texttt{PYTHIA} and \texttt{WHIZARD}. The signal process is defined in \texttt{WHIZARD} using built-in models THDM or MSSM (both models give identical cross sections) and the Higgs boson masses are set in the \texttt{SINDARIN} command files. Using \texttt{PYTHIA} for signal production requires mass spectrum files in \texttt{LHA} format \cite{LHA} which are generated using \texttt{2HDMC}. Tables \ref{Xsig} and \ref{Xbag} show the signal and background cross sections respectively with values from the two generators. Final results are normalized to \texttt{WHIZARD} cross sections. The $Z^{(*)}/\gamma^*$ has been generated in the fully hadronic final state and is slightly higher than the corresponding result from \texttt{PYTHIA}. The $ZZ$ background includes only $Z$ boson pair production. The \texttt{WHIZARD} cross sections include beam spectrum and ISR.
 \begin{table}[h]
\centering
\begin{tabular}{|ccccc|}
	\hline
	\multicolumn{5}{|c|}{Signal process}\\
	\hline
     Benchmark point & BP1 & BP2 & BP3 & BP4 \\
	\hline
	W: Total $\sigma [fb]$ & 12.3 & 9.1 & 8.0 & 5.9 \\
	\hline
	P: Total $\sigma [fb]$ & 12.1 & 9.4 & 8.5 & 6.2 \\
	\hline
	BR$(H \to b\bar{b})$ & 0.71 & 0.62 & 0.51 & 0.38 \\
	\hline
	BR$(A \to b\bar{b})$ & 0.54 & 0.13 & 0.29 & 0.16 \\
	\hline
	 $\sigma \times BR~ [fb]$ & 4.7 & 0.73 & 1.18 & 0.36  \\
	\hline
\end{tabular}
\caption{Signal cross sections assuming different benchmark points. The letters ''W`` and ''P`` denote the results from \texttt{WHIZARD} and \texttt{PYTHIA}.}
\label{Xsig}
\end{table}

 \begin{table}[h!]
	\centering
\begin{tabular}{|cccc|}
	\hline
	\multicolumn{4}{|c|}{Background processes}\\
	\hline
 & $ZZ$ & $Z/\gamma^*$ &$t\bar{t}$ \\
	\hline
	W: $\sigma[fb]$& ${181}$ & ${3473}$& ${197}$ \\
	\hline
	P: $\sigma[fb]$& ${176}$ & ${3015}$& ${211}$ \\
	\hline
\end{tabular}
\caption{Background cross sections from \texttt{WHIZARD} (denoted by ''W``) and \texttt{PYTHIA} (''P``).}
\label{Xbag}
\end{table}

\subsection{$b$-tagging} 
Since there are four $b$-jets in the signal final state, the $b$-tagging algorithm is used based on MC truth matching using ILCgen card which contains $p_T,\eta$ dependent selection efficiencies for $b$-jets, $c$-jets and light jets ($u, d, s$). There are three benchmark scenarios for the average $b$-tagging efficiency: 80$\%$, 70$\%$ and 50$\%$.     

Every signal or background event is required to contain exactly four $b$-jets. Therefore, the event selection starts with selecting events with exactly four jets requiring all of them to pass the $b$-tagging. 

The kinematic requirement for the jet selection is $E_T>10$ GeV (soft jet veto) and $|\eta|<2$ (central jet selection).

\begin{figure}[h]
	\centering
\includegraphics[width=0.49\textwidth]{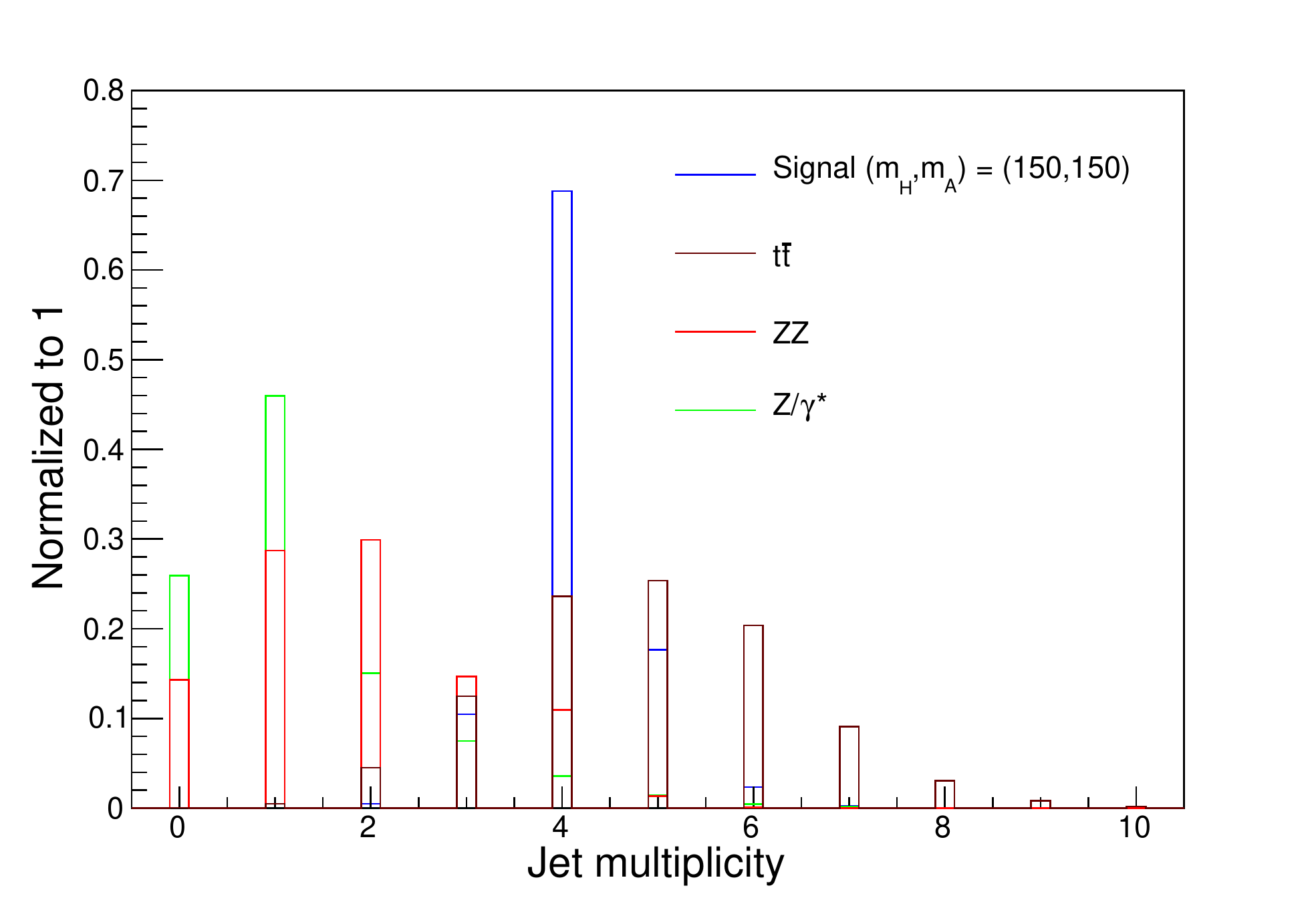}  
	\caption{The reconstructed jet multiplicity in signal and background events. Only BP1 has been shown representing the signal. \label{jetmul}}
\end{figure}

\begin{figure*}[h!]
	\centering
    \subfigure[]{\includegraphics[width=0.32\textwidth,height=0.2\textheight]{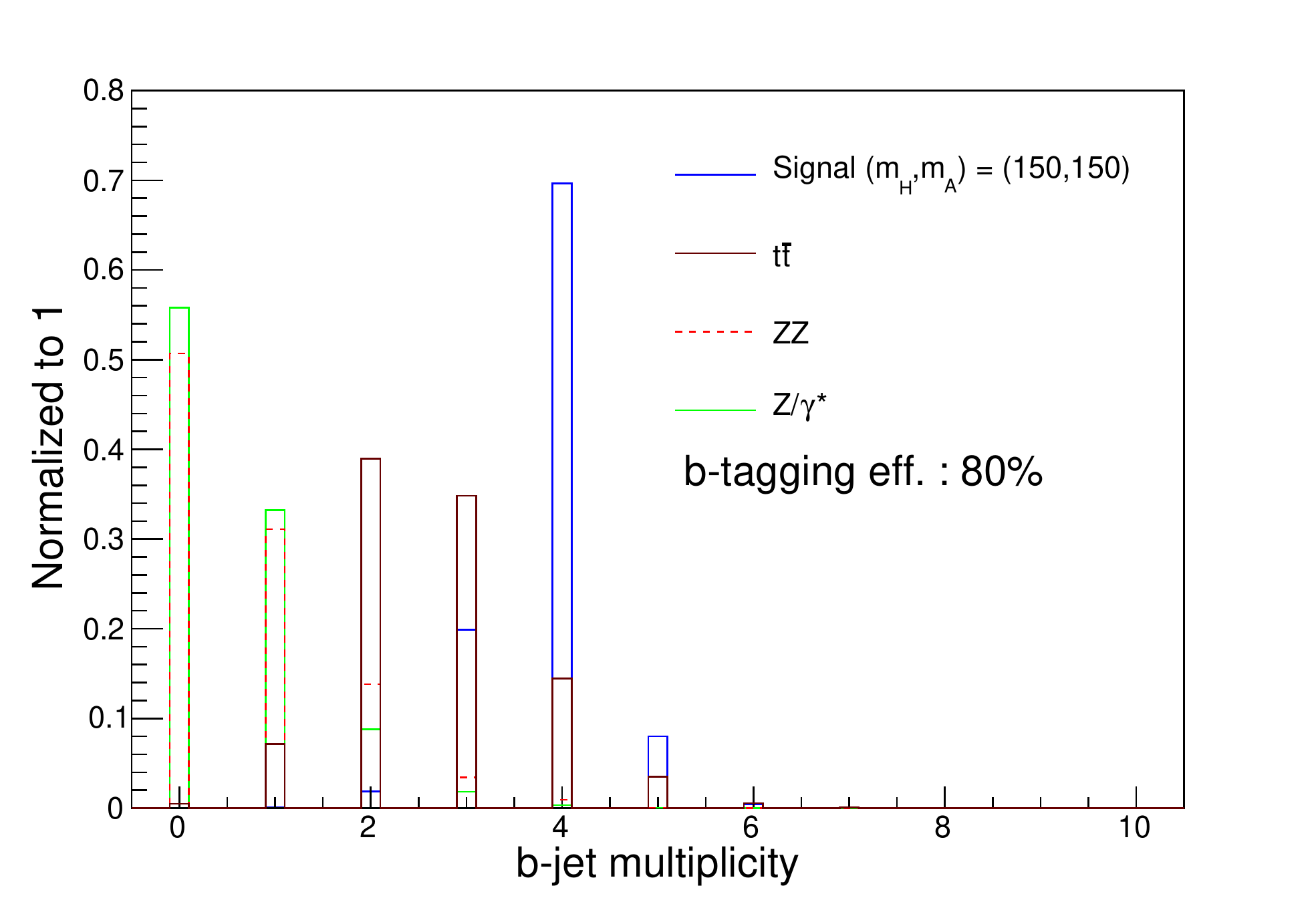}}  
    \subfigure[]{\includegraphics[width=0.32\textwidth,height=0.2\textheight]{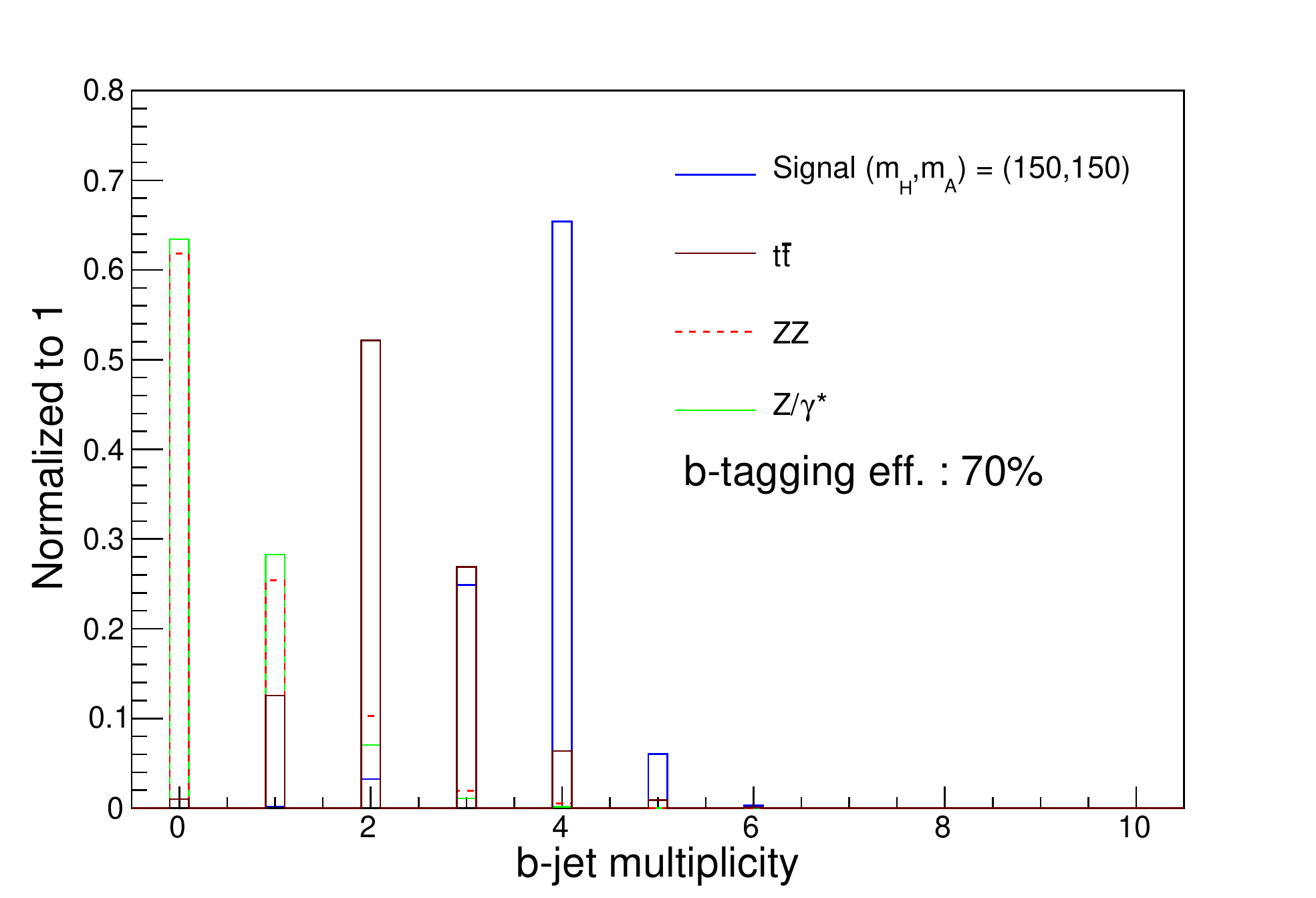}}
    \subfigure[]{\includegraphics[width=0.32\textwidth,height=0.2\textheight]{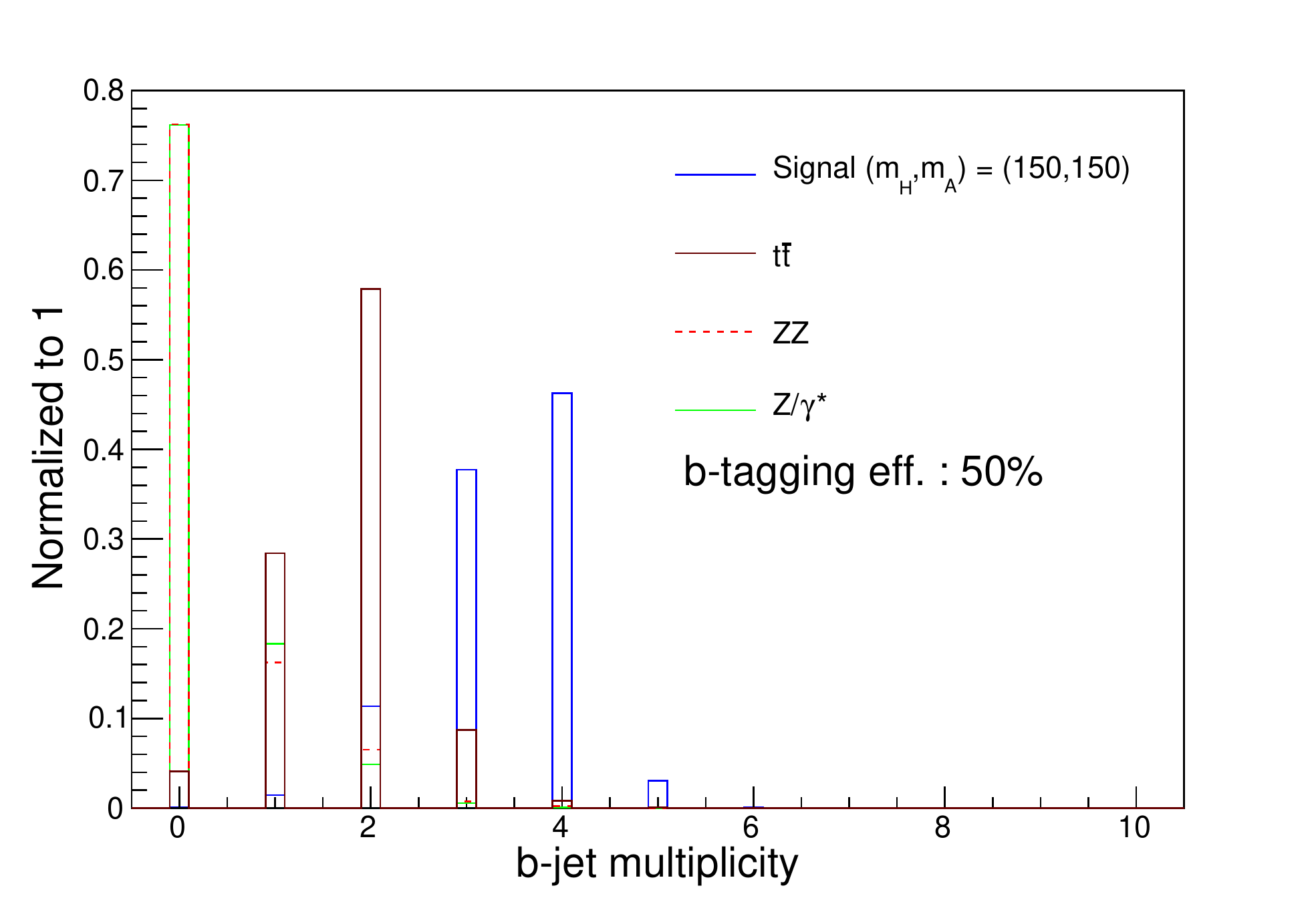}}
	\caption{The $b$-jet multiplicity in signal and background events with different $b$-tagging efficiencies.} 
	\label{bjetmul}
\end{figure*}

Figure \ref{jetmul} shows the jet multiplicity in signal and background events and provides the reason for excluding events with more than four jets which are dominated by $t\bar{t}$. Figure \ref{bjetmul} shows the $b$-jet multiplicities in three $b$-tagging scenarios. Although choosing tighter $b$-tagging scenario reduces the four $b$-jet selection efficiency in signal events from 70$\%$ to roughly 50$\%$, the contribution of $t\bar{t}$ events in the four $b$-jet bin is the most important point due to the very high cross section of this process. However, as is seen in Fig. \ref{bjetmul}, the contribution of this background in the  four $4$-jet bin is suppressed very well by choosing the third $b$-tagging scenario with average efficiency of 50$\%$.

\subsection{Kinematic correction}
When an event containing four $b$-jets is selected, a kinematic correction is applied on the $b$-jets four momenta according to what is expected from momentum and energy conservation. 

The energy conservation relies on the fact that at lepton colliders, the beam energy is known within the uncertainty arised from the ISR and beamstrahlung. At hadron colliders, the situation is more complicated due to the fact that the effective center of mass energy varies event by event due to the parton distribution functions. The beam spectrum and ISR certainly affect the correction procedure in this analysis, however, as can be seen from the final results, a reasonable performance is observed even including such effects.

It should be noted that the beam crossing angle can also affect the kinematic correction performance which implies that there is no total momentum component in any direction. The effect of beam crossing angle can easily be activated in \texttt{WHIZARD}. However, since it is not yet implemented in \texttt{DELPHES}, we did not apply it for the current analysis.   

The set of four equations representing four-momentum conservation includes four correction factors which are named $c_i$ with $i \in 1-4$ assigned to the four $b$-jets in the event:
\begin{eqnarray}\label{conservation}
c_1p^{b_1}_x+c_2p^{b_2}_x+c_3p^{b_3}_x+c_4p^{b_4}_x&=&0\nonumber\\
c_1p^{b_1}_y+c_2p^{b_2}_y+c_3p^{b_3}_y+c_4p^{b_4}_y&=&0\nonumber\\
c_1p^{b_1}_z+c_2p^{b_2}_z+c_3p^{b_3}_z+c_4p^{b_4}_z&=&0\nonumber\\
c_1E^{b_1}   +c_2E^{b_2}   +c_3E^{b_3}   +c_4E^{b_4}   &=&\sqrt{s}.\nonumber\\
\end{eqnarray}
Therefore all four momentum components of each $b$-jet receive the same correction factor without changing the $b$-jet flight direction. In order to avoid negative energy values, all correction factors are required to be positive. 

The set of Eq. \ref{conservation} consists of four linear equations with four unknowns. The momentum components of the four $b$-jets make the $4\times 4$ coefficient matrix which has a non-zero determinant due to the random nature of the momentum components in events. Therefore, a unique non-trivial solution for the four coefficient factors is expected.

The energy correction may change the order of the $b$-jets in the list as they are sorted according to descending energies by default. Therefore, the energy sorting algorithm is applied again after the correction and the four $b$-jets are selected for the rest of the analysis. 

Tables \ref{preselection_s} and \ref{preselection_b} present the pre-selection efficiencies of the jet reconstruction and four jet selection, $b$-tagging (having four $b$-tagged jets in the event) and the positive correction factor requirement in signal and background processes.

According to Tab. \ref{preselection_s}, the correction efficiencies increase with increasing the Higgs boson masses. In other words, the correction performance is better for harder jets from heavier Higgs bosons and reaches 86$\%$ for BP4. 

\begin{figure}[h!]
	\centering
     \includegraphics[width=0.49\textwidth]{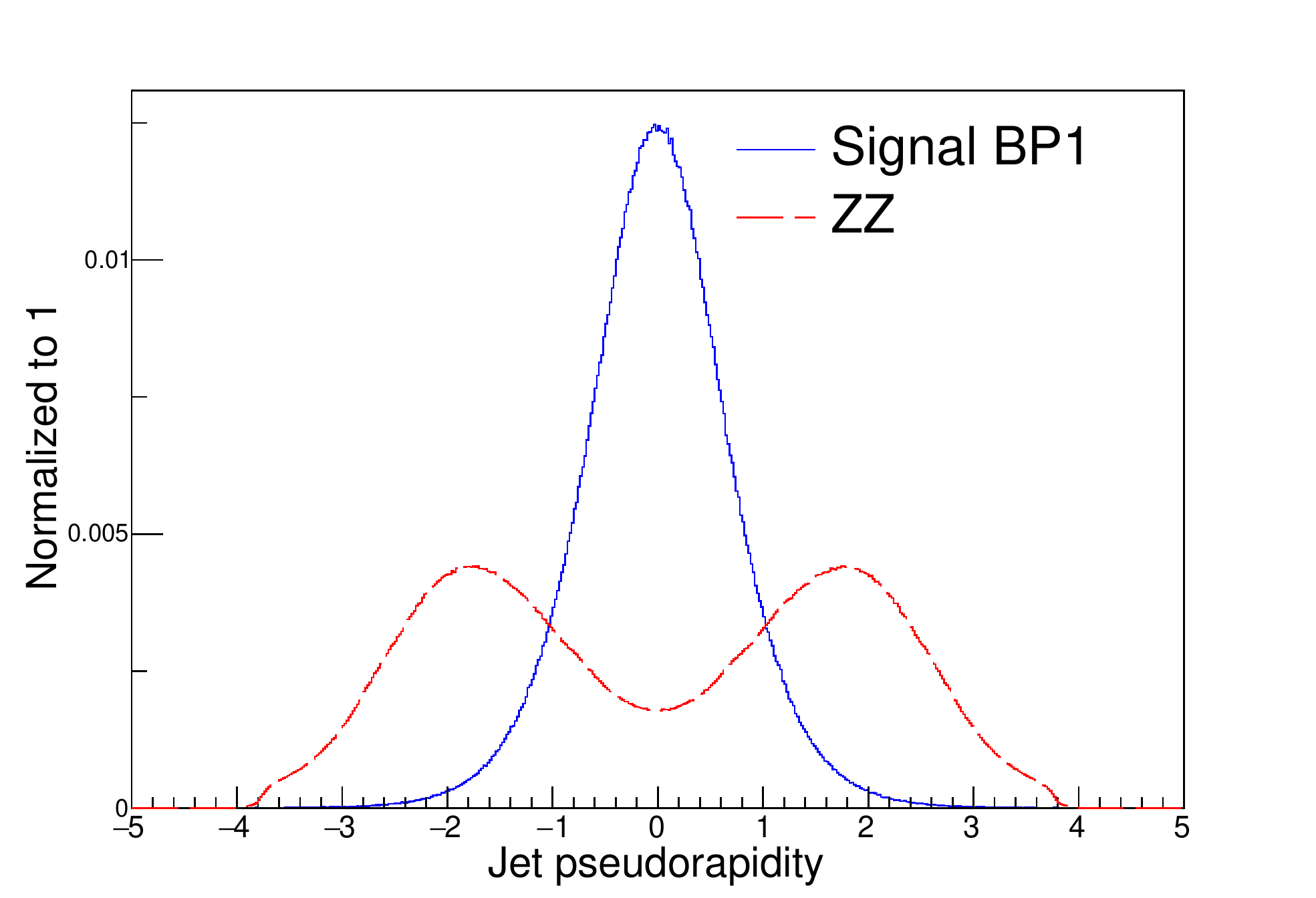}
	\caption{Reconstructed jet pseudorapidity in signal (BP1) and $ZZ$ background.}
	\label{eta}
\end{figure}

The single $Z/\gamma^*$ production is expected to have a low four jet selection efficiency. However, as seen from Tab. \ref{preselection_b}, The $ZZ$ background is also suppressed very well by the four jet selection which is due to the performance of the kinematic cuts applied on the jets. The main kinematic difference between the signal and $ZZ$ events is due to the pseudorapidity distributions which are shown in Fig. \ref{eta}. 

As seen from Fig. \ref{eta}, the jets from $ZZ$ events tend to proceed through the forward/backward region resulting in central jet selection efficiency ($|\eta|<2$) of about 66$\%$. Therefore, all four jets appear in $|\eta|<2$ region with probability of $\sim 18\%$. The cut on the jet transverse energy ($E_{\textnormal{jet}}\times \sin\theta$) further reduces the selection efficiency to 11$\%$ which appears in Tab. \ref{preselection_b}. The $b$-tagging requirement suppresses this background further, which, followed by the correction efficiency, results in preselection efficiency of $10^{-3}$. 

\begin{figure}[h!]
	\centering
     \includegraphics[width=0.49\textwidth]{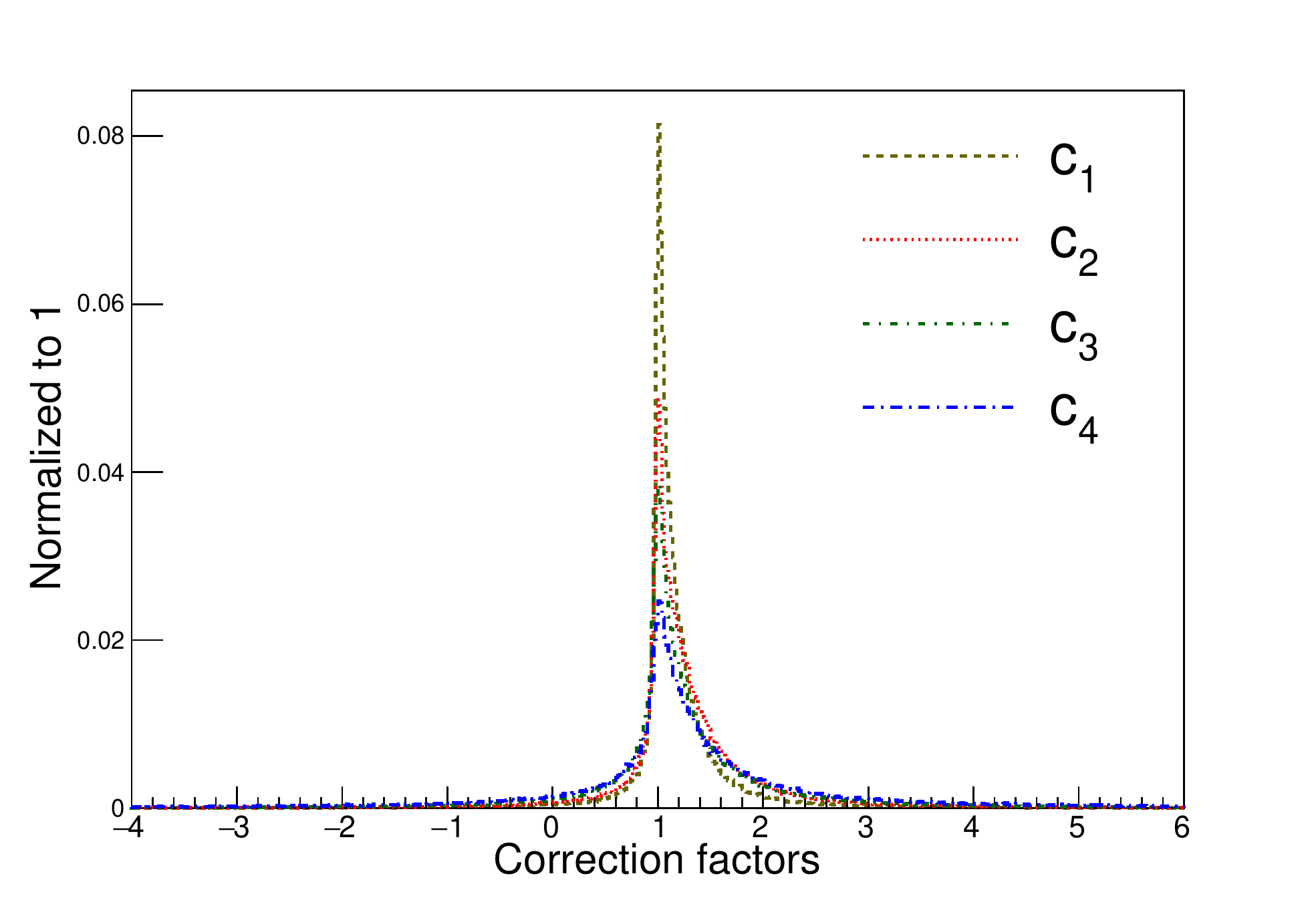}
	\caption{Correction factor distributions (BP1).}
	\label{cfactors}
\end{figure}

Figure \ref{cfactors} shows correction factor distributions in signal events (BP1) with average values of 1.16, 1.25, 1.21 and 1.26. While the mean values of the distributions are close to unity, their widths (RMS values) are 0.6, 1.1, 1.45, 1.98 for $c_1$ to $c_4$ respectively and again shows better performance of the correction for harder jets.

In order to verify the correction efficiency, using Fig. \ref{cfactors} as the example, the four correction factors $c_1$ to $c_4$ are found to be positive with efficiencies of 98, 96, 92 and 85$\%$ respectively. Since all factors are required to be positive, the quoted efficiencies are multiplied to yield a total efficiency of $\sim70\%$ which is what we obtain in the event analysis.

\begin{figure}[h!]
	\centering
     \includegraphics[width=0.49\textwidth]{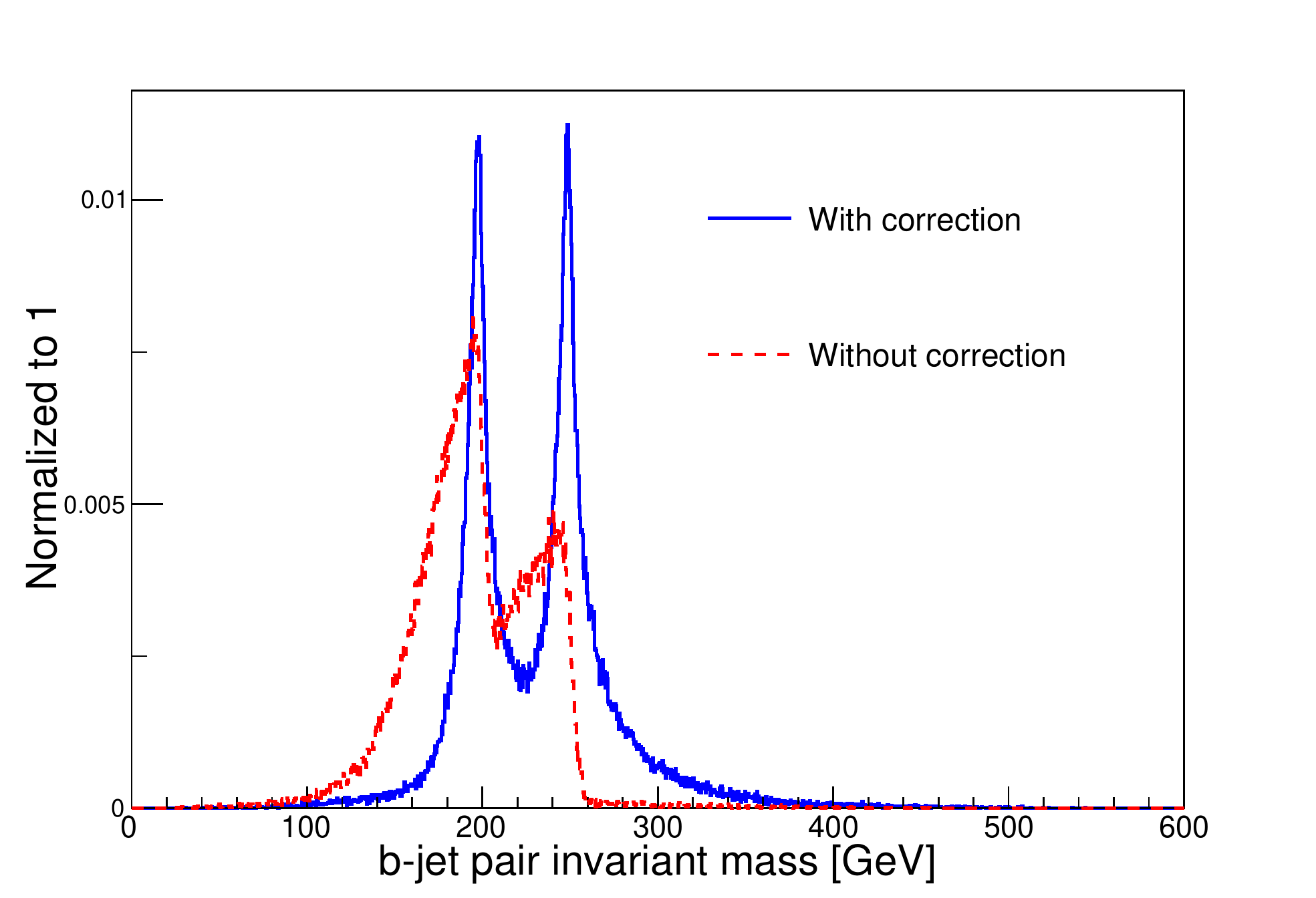}
	\caption{The corrected and uncorrected distributions of the $b$-jet pair invariant mass (BP2).}
	\label{diff}
\end{figure}

The effect of the kinematic correction on the $b$-jet pair invariant mass is shown in Fig. \ref{diff} using BP2 as the example. Details of the $b$-jet pair selection are presented in the next sections. 

The correction factor sensitivity to the reconstructed jet energies is verified by estimating the uncertainty of the correction factors due to the jet energy smearing. 

\begin{figure}[h!]
	\centering
     \includegraphics[width=0.49\textwidth]{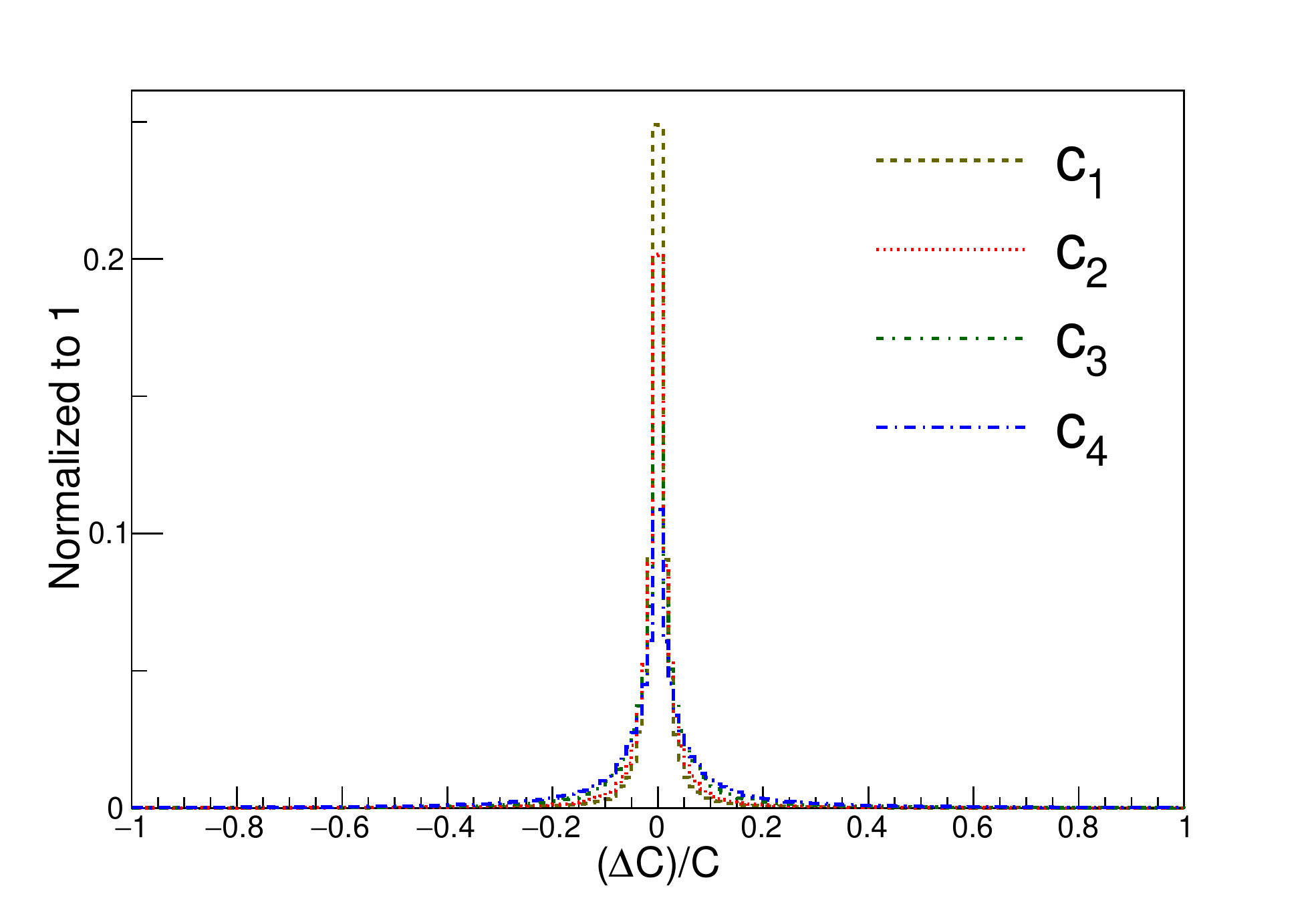}
	\caption{The correction factor uncertainty due to 1$\%$ smearing on the jet energies.}
	\label{cerror}
\end{figure}

In order to do so, $1\%$ additional smearing is applied on the jet four-momentum components on event-by-event basis and the distributions of relative errors of the correction factors are obtained. Results are shown in Fig. \ref{cerror} and can be regarded as the correction factor smearing due to $1\%$ uncertainty in the jet energies. The average uncertainties are 9, 11, 14 and 18$\%$ for $c_1$ to $c_4$ respectively. 

The above estimates are one of the sources of the total uncertainty in the final distributions. However, a detailed study of different sources of uncertainties and their influence on the final distributions is beyond the scope of the current analysis and can be performed in a more detailed analysis based on full simulation of the detector.
\begin{table}[h]
	\centering
	\begin{tabular}{|ccccc|}
		\hline
		&$BP1$ & $BP2$ & $BP3$ & $BP4$ \\
		\hline
		Four jet eff. & 0.69 & 0.47 & 0.41 & 0.34\\
		\hline
		Four $b$-jet eff. & 0.54 & 0.53 & 0.52 & 0.51\\
		\hline
		Correction eff. &  0.71 & 0.80 & 0.82 & 0.86 \\
		\hline
		Total pre-sel. eff. & 0.26 & 0.20 & 0.18 & 0.15 \\
		\hline
	\end{tabular}
	\caption{Preselection efficiencies of the signal events. }
	\label{preselection_s}
\end{table}
\begin{table}[h]
	\centering
	\begin{tabular}{|cccc|}
		\hline
 & $ZZ$ & $Z/\gamma^*$ &$t\bar{t}$ \\
		\hline
		Four jet eff. & 0.11 & 0.035 & 0.24\\
		\hline
		Four $b$-jet eff. & 0.18 & 0.004 & 0.002\\
		\hline
		Correction eff. &  0.54 & 0.54 & 0.31\\
		\hline
		Total pre-sel. eff. &  1$\times 10^{-3}$ & 7$\times 10^{-5}$ & 2$\times 10^{-4}$\\
		\hline
	\end{tabular}
	\caption{Preselection efficiencies of the background events. }
	\label{preselection_b}
\end{table}

\subsection{$b$-jet pair selection based on their energies}\label{bb14}
Finding the true $b$-jet combination for Higgs boson reconstruction relies on two approaches. In the first approach, we note that decay products which move closer to the beam axis in the Higgs boson rest frame, acquire the highest and lowest energies when the Lorentz boost is applied to transform them from the Higgs boson frame to the laboratory frame. The other decay products belong to the latter Higgs boson whose decay occurs at a larger angle with respect to the beam axis. Therefore having sorted the four $b$-jets in terms of their energies, $b_1$ and $b_4$ are expected to be the decay products of one Higgs boson and $b_2$ and $b_3$ from the other. 

\begin{figure}[h!]
	\centering
     \includegraphics[width=0.49\textwidth]{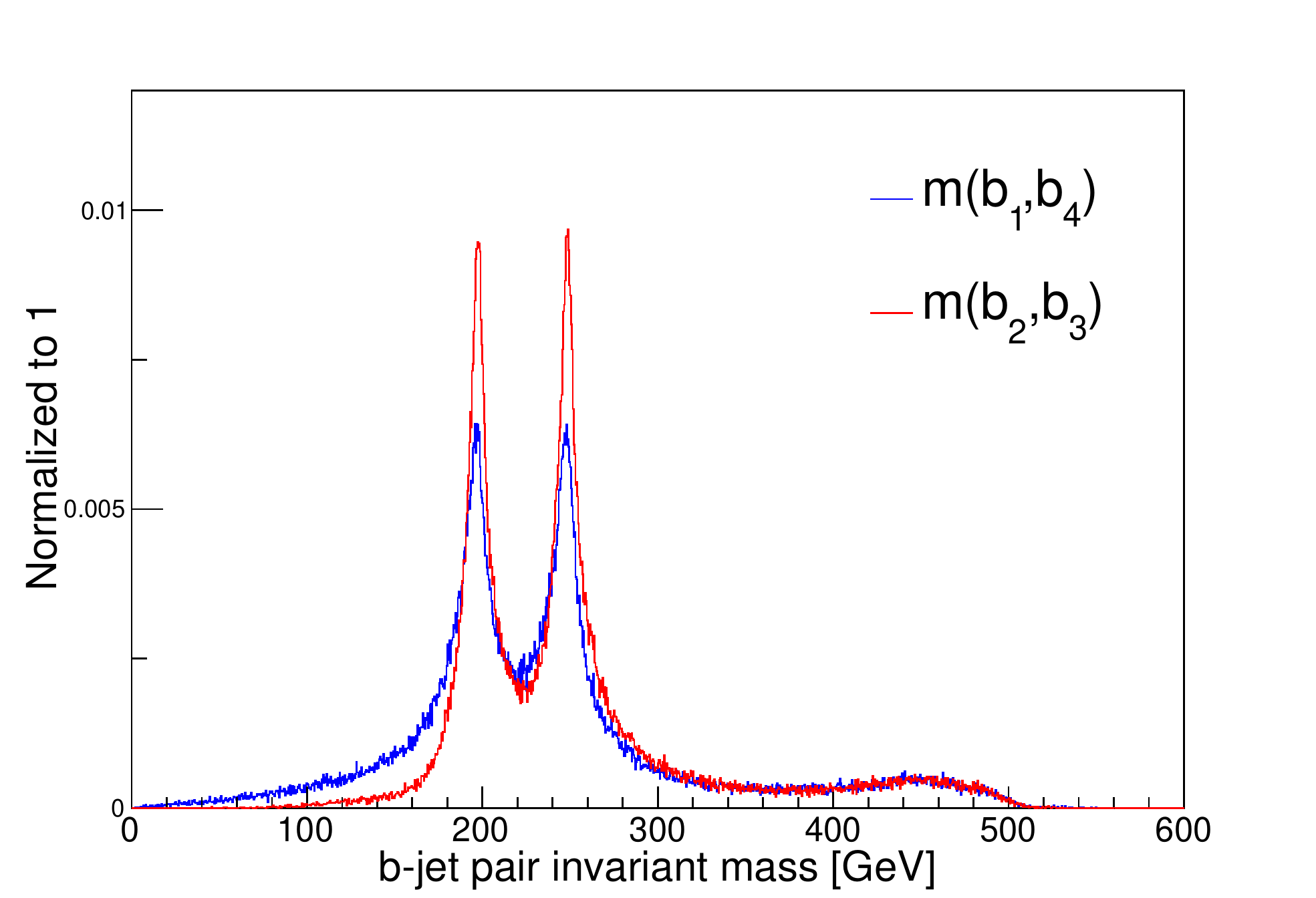}
	\caption{Comparison between the two selection scenarios: $b_1b_4$ vs $b_2b_3$}
	\label{bb1423}
\end{figure}

Since the decay products of the Higgs bosons fly at random angles with respect to the beam in each event, two possible scenarios may occur: $H \to b_1 b_4$, $A \to b_2 b_3$ or $H \to b_2 b_3$, $A \to b_1 b_4$. Some events choose the former scenario and the others choose the latter. Therefore, the $b$-jet pair invariant mass distribution dramatically shows both Higgs boson signals even if the distribution is obtained using only $b_1b_4$ or $b_2b_3$. If the Higgs boson masses are different enough to distinguish their signals, two separated peaks are observed, otherwise only one peak is observed. Figure \ref{bb1423} shows an example of the signal distributions (BP2) with two pairings, i.e., $b_1b_2$ and $b_2b_3$. The latter pairing results in slightly better distribution due to using more central $b$-jets.   

\subsection{$b$-jet pair selection based on their spatial distance}\label{DR4}
In the alternative approach, the two $b$-jet pairs are selected requiring minimum $\Delta R$ between each pair. In order to do so, sum of the two $\Delta R$ values are required to be minimum for the selected pairs.

\begin{figure}[h!]
	\centering
     \includegraphics[width=0.49\textwidth]{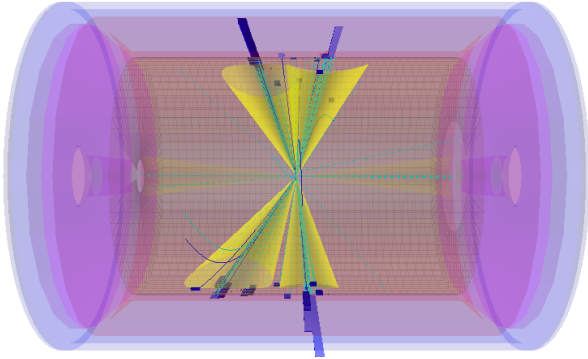}
	\caption{3D view of a signal event in the four b-jet final state in the detector. Jets from each Higgs boson tend to be collinear. The visible non-zero total momentum along the beam axis is due to the ISR+beamstrahlung as verified by inspecting the \texttt{WHIZARD} event file which was used for the simulation.}
	\label{3D}
\end{figure}

The idea is based on the fact that, in general, $b$-jets from the decay of a particle, are expected to proceed at smaller $\Delta R$ values compared to two randomly selected $b$-jets, each one belonging to a different particle. Figure \ref{3D} shows an example of a signal event (BP1) in four $b$-jet final state in the detector using DELPHES Display module \cite{DELPHES}.  

The above requirement turns out to perform the $b$-jet pairing very similar to the previous scenario. As an example, in 97$\%$ of the BP1 events, the selected pairs are $b_1b_4$ and $b_2b_3$ and $\Delta R(b_1b_4)+\Delta R(b_2b_3)$ is minimum among other possible combinations. 

Therefore, for the final event selection, the two approaches described in sub-sections \ref{bb14} and \ref{DR4} are combined by requiring $\min (\Delta R(b_ib_j)+\Delta R(b_kb_l))$ and then demanding $i=1,~j=4,~k=2~\textnormal{and}~l=3$. 

 The performance of this requirement depends on the Higgs boson masses and their momenta and decreases when $m_H+m_A$ reaches the kinematic threshold $\sqrt{s}$. In such cases, the two Higgs bosons are almost created at rest with their decay products flying back-to-back at the maximum $\Delta R$ in the laboratory frame. However, we keep this requirement to suppress the large $Z/\gamma^*$ background, which otherwise, extends to the signal region. 

\begin{figure}[h!]
	\centering
     \includegraphics[width=0.49\textwidth]{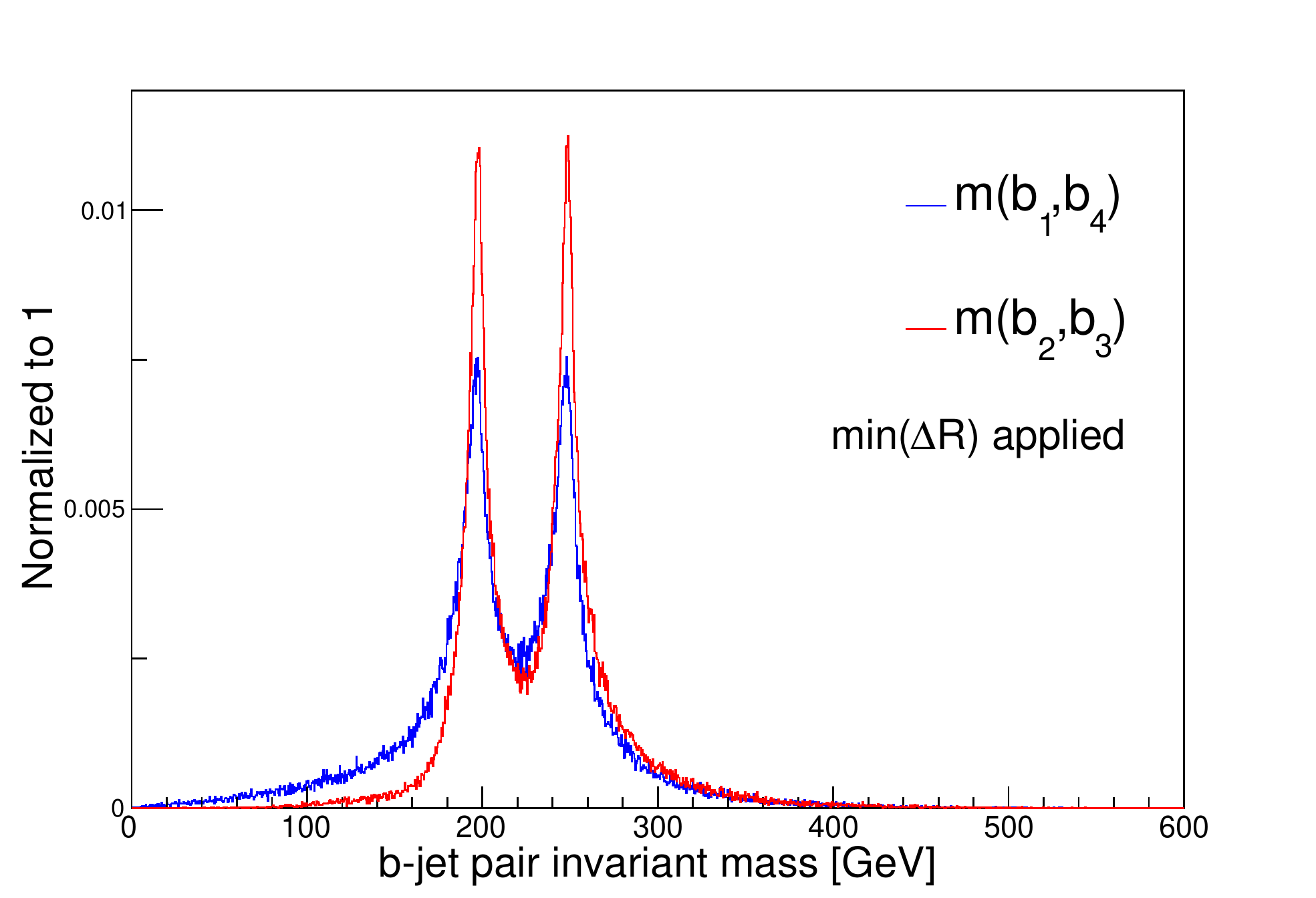}
	\caption{Comparison between the two selection scenarios: $b_1b_4$ vs $b_2b_3$ with min($\Delta R$) requirement applied.}
	\label{bb1423DR}
\end{figure}

Figure \ref{bb1423DR} compares the two pairing scenarios with min($\Delta R$) applied. The two distributions are in general better than those shown in Fig. \ref{bb1423} and the high energy tail is well suppressed. Again the $b_2b_3$ pairing shows a better distribution compared to $b_1b_4$. Therefore the final event distributions are obtained using $b_2b_3$ with min($\Delta R$) requirement applied which is the best scenario among the four possible choices shown in Figs. \ref{bb1423} and \ref{bb1423DR}. 

\begin{figure*}[h!]
	\centering
    \subfigure[]{\includegraphics[width=0.32\textwidth,height=0.2\textheight]{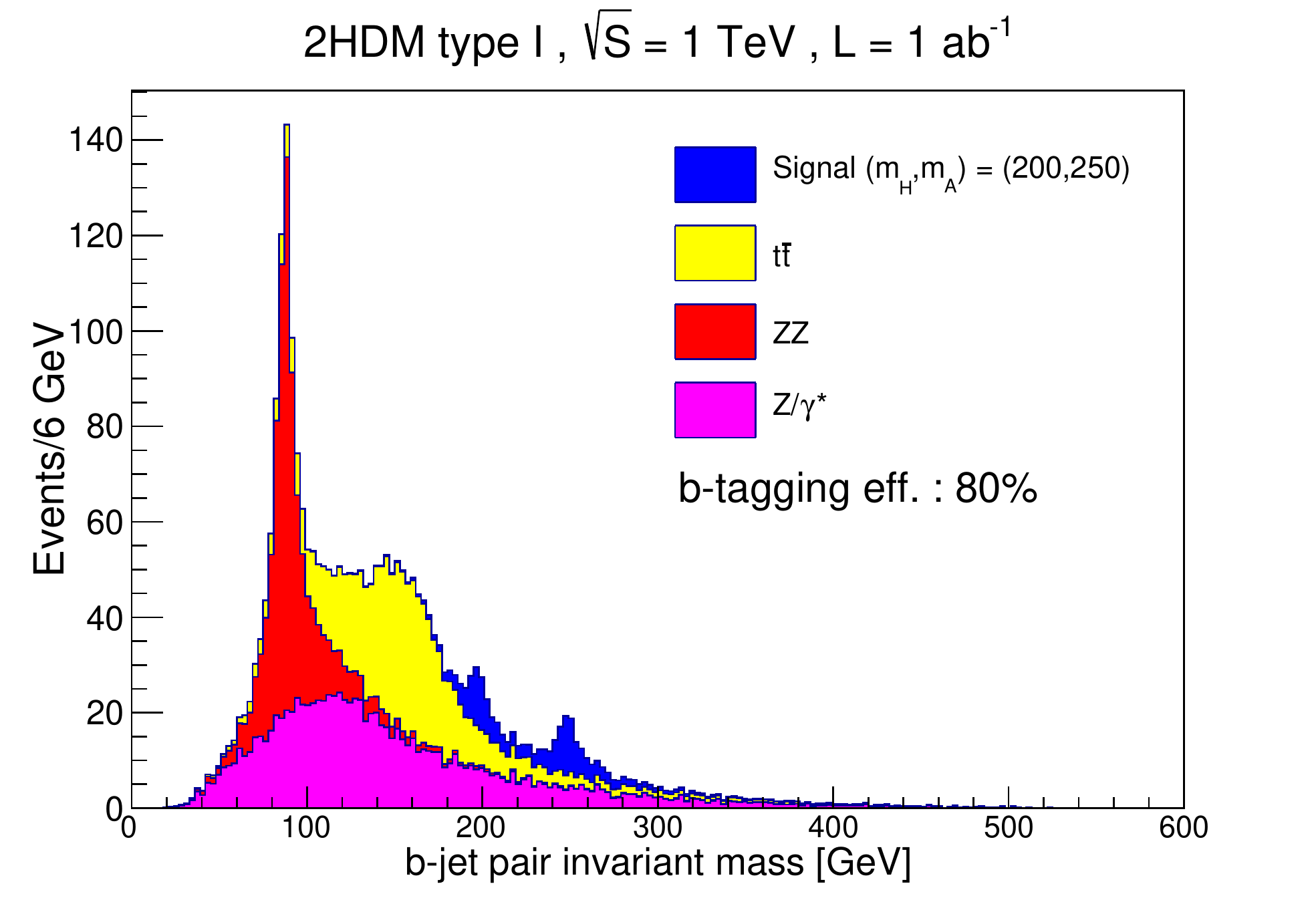}}  
    \subfigure[]{\includegraphics[width=0.32\textwidth,height=0.2\textheight]{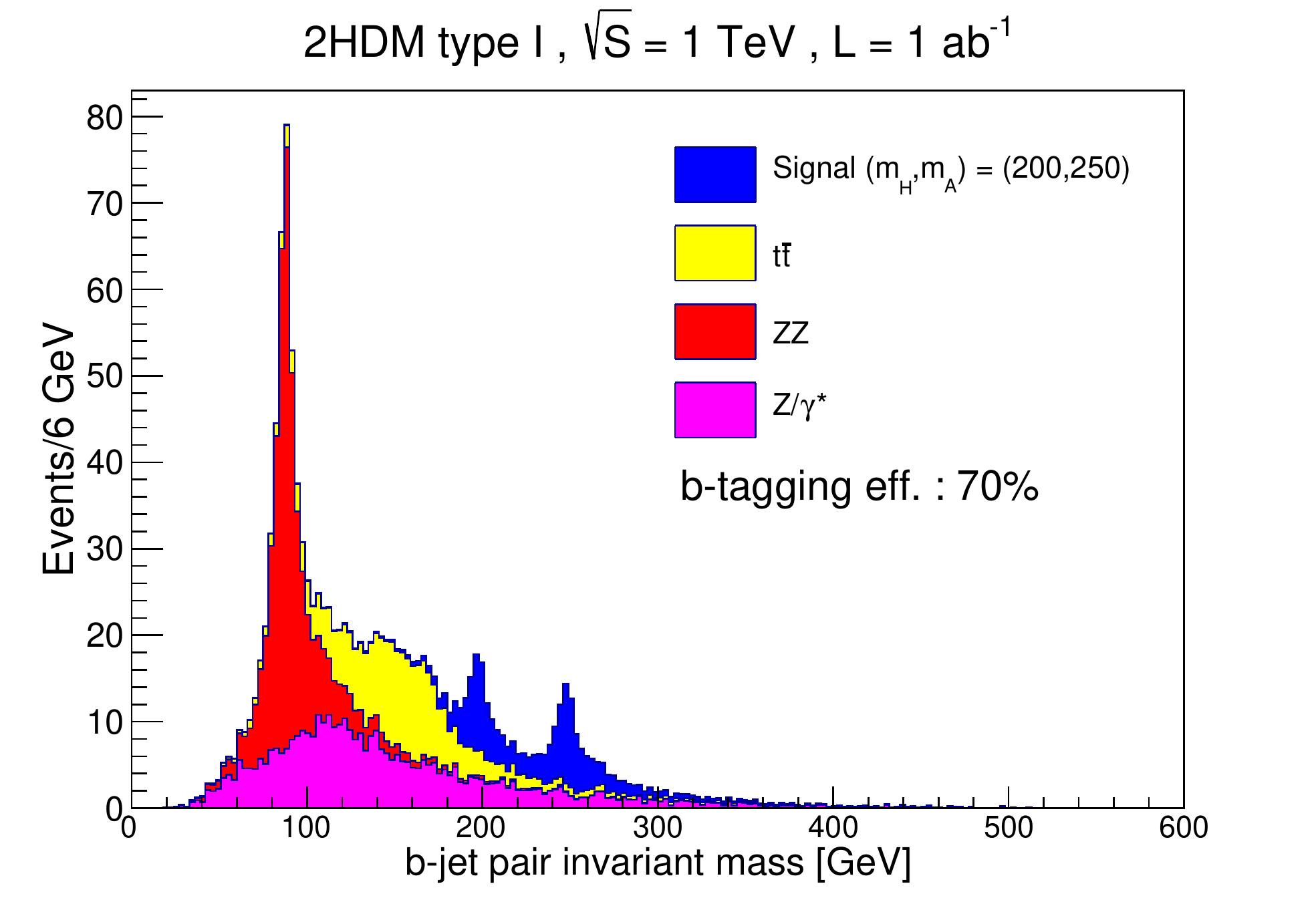}}
	\caption{The $b$-jet pair invariant mass distributions in signal and background events with $b$-tagging efficiencies of 80$\%$ and 70$\%$ respectively. }
	\label{bb8070}
\end{figure*}
\begin{figure*}[h!]
	\centering
    \subfigure[]{\includegraphics[scale=0.4]{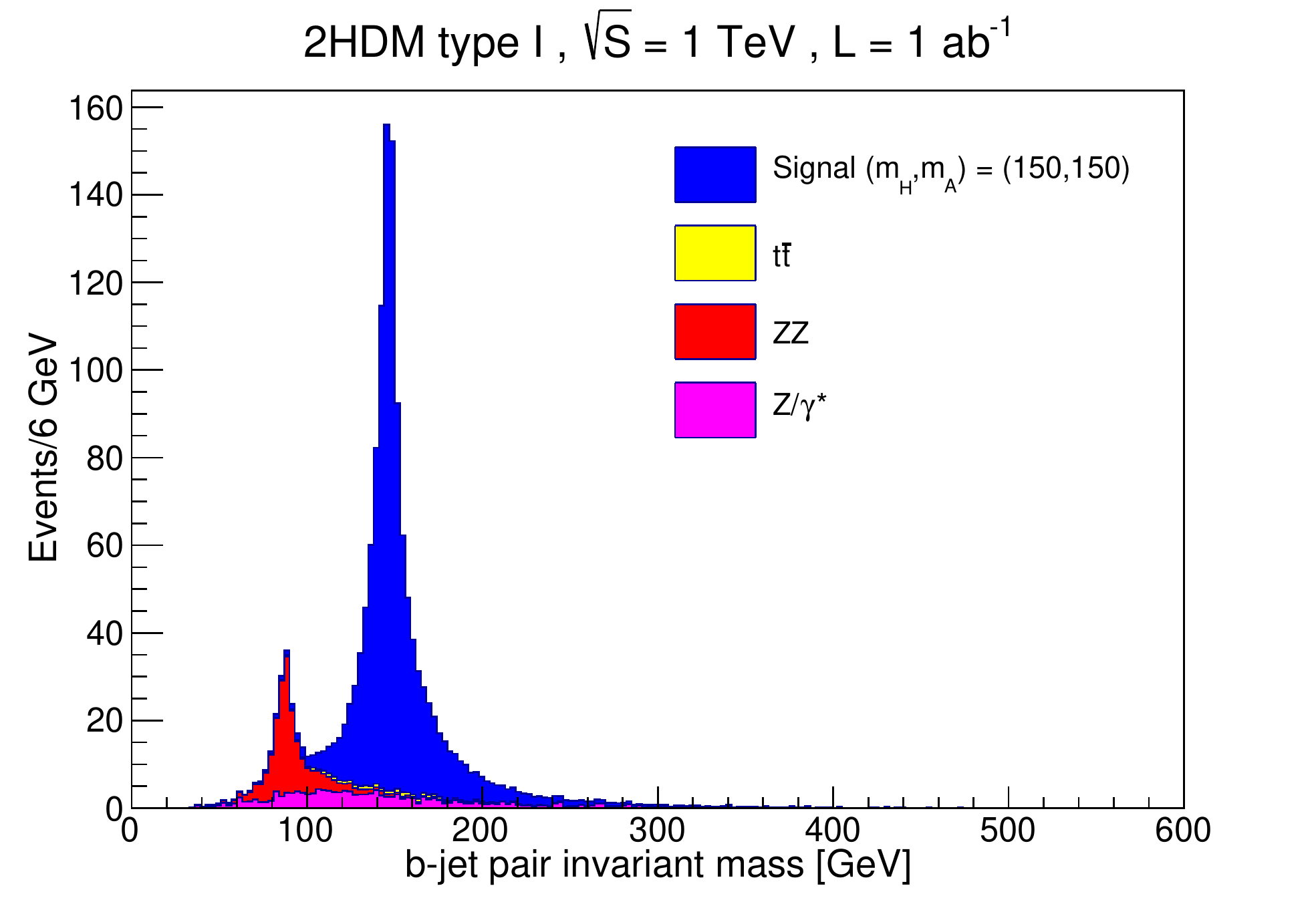}}  
    \subfigure[]{\includegraphics[scale=0.4]{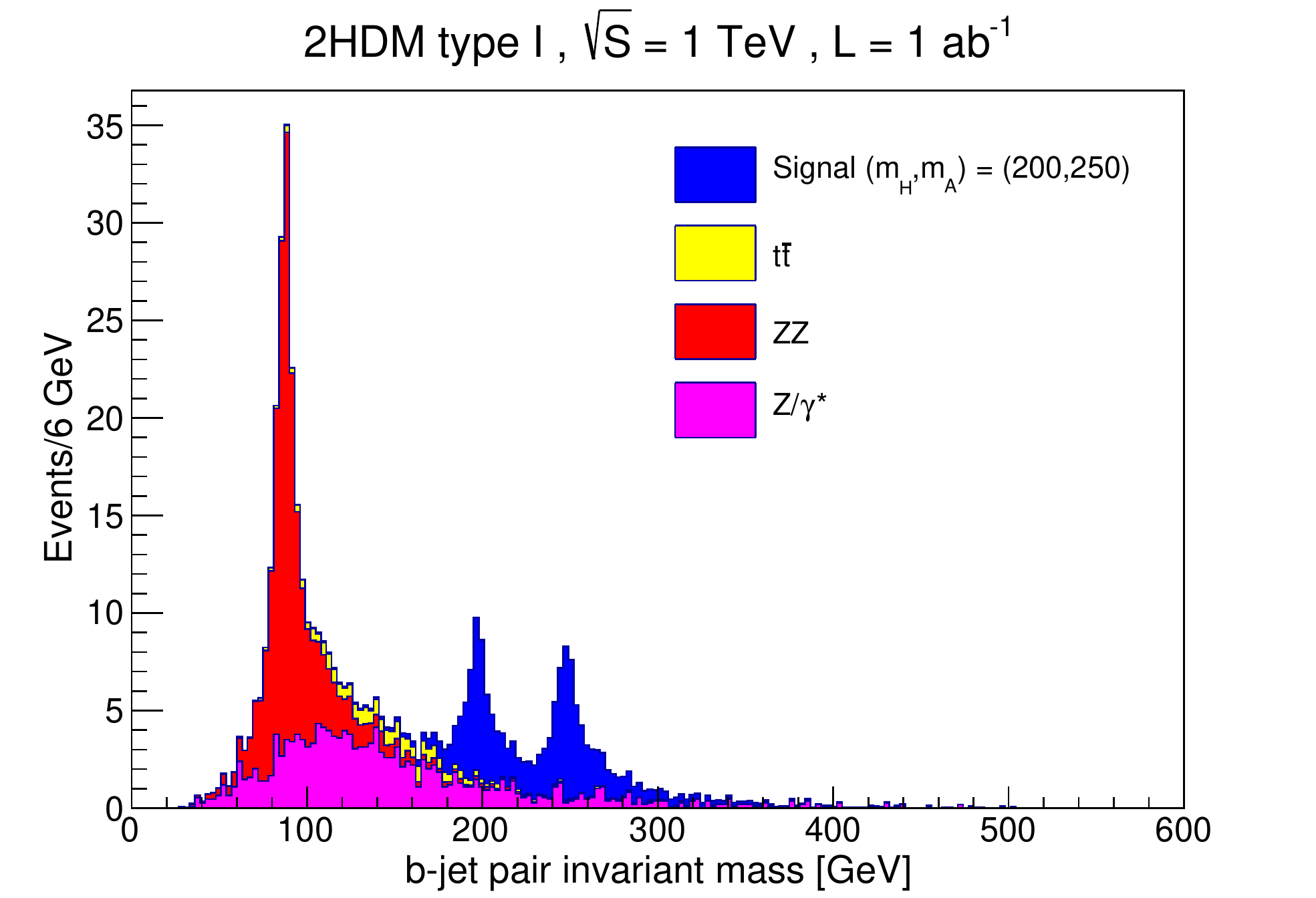}}
    \subfigure[]{\includegraphics[scale=0.4]{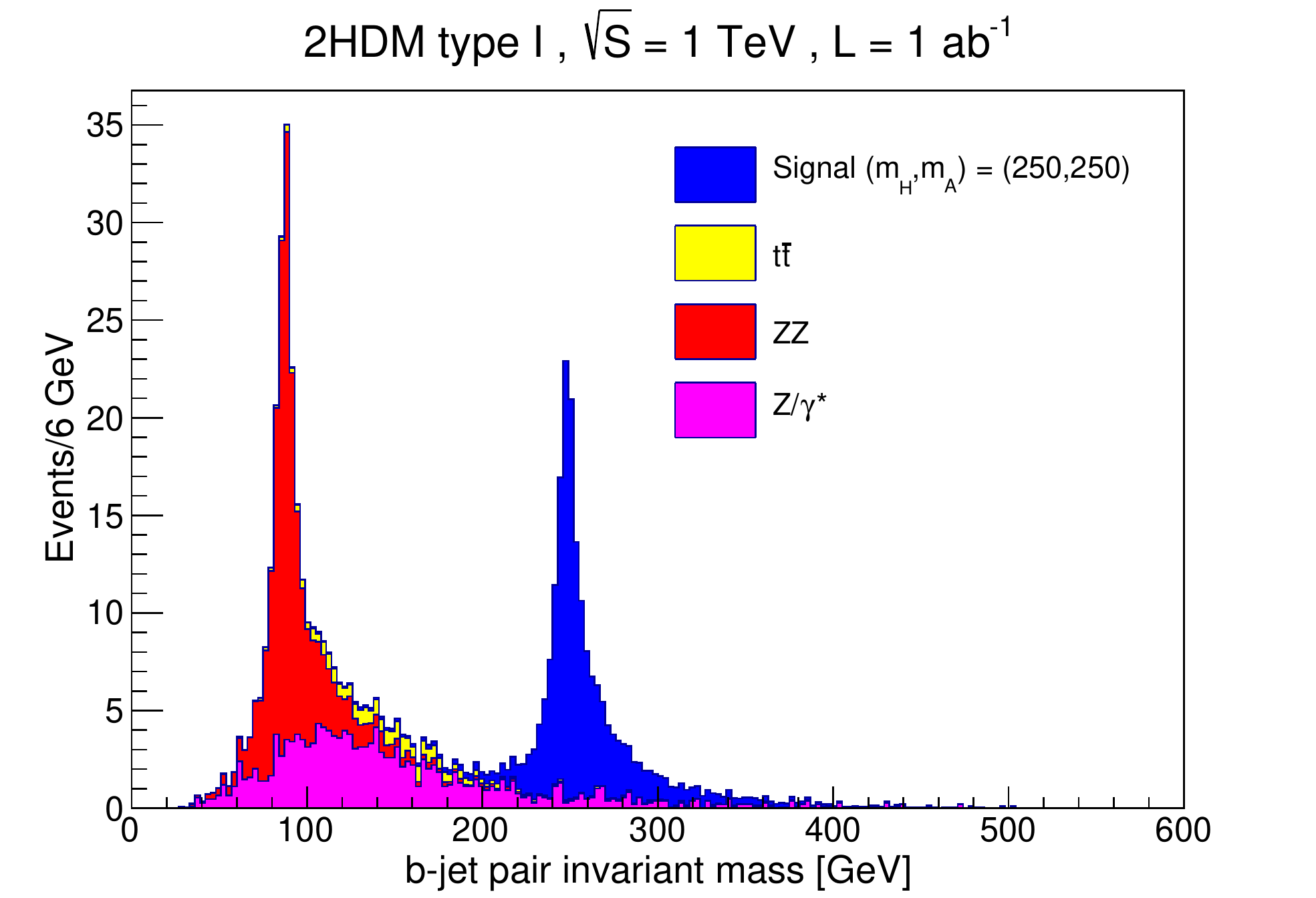}}
     \subfigure[]{\includegraphics[scale=0.4]{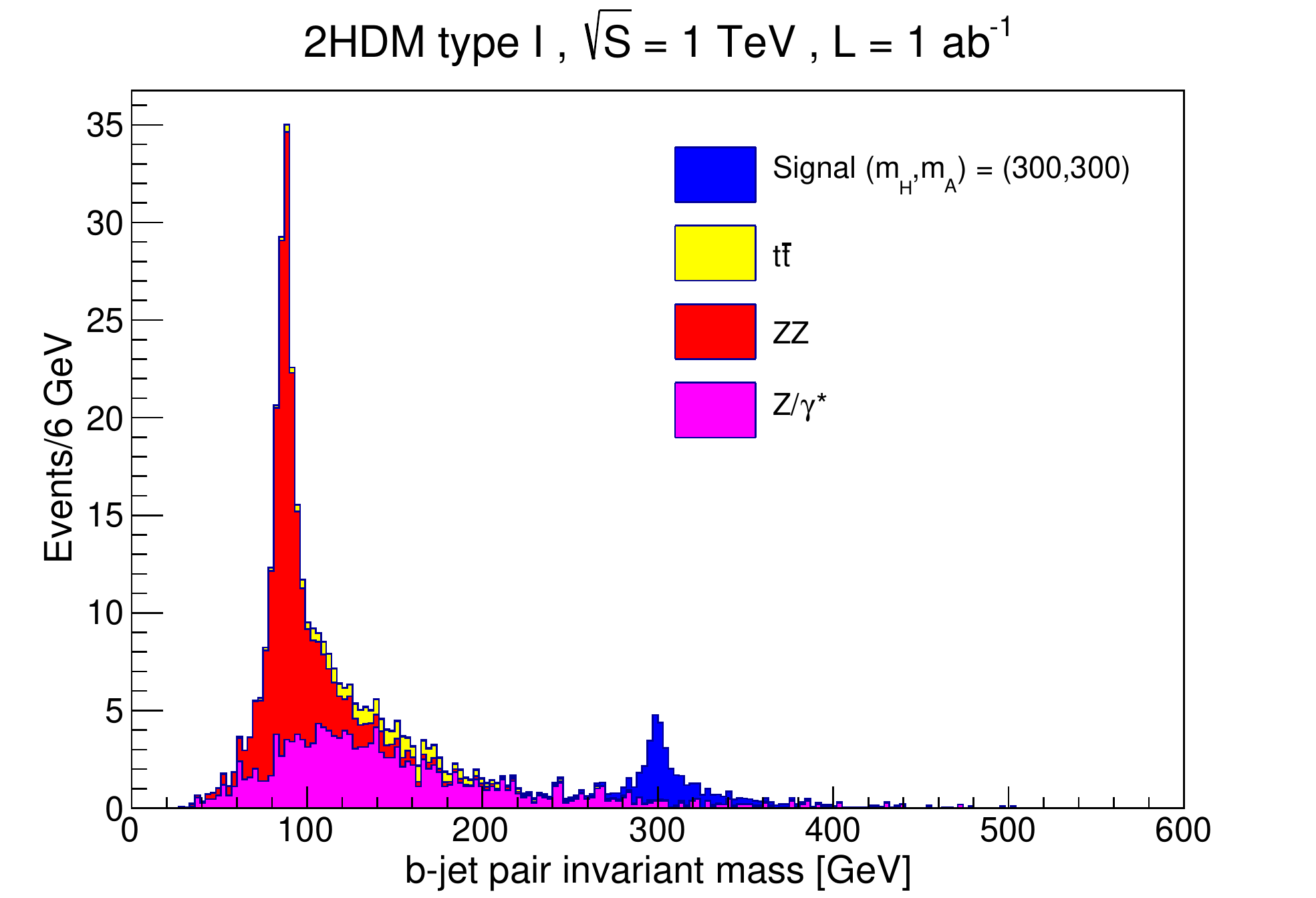}}
	\caption{The $b_2b_3$ pair invariant mass distributions in signal and background events at $\sqrt{s}=1$ TeV. The min($\Delta R$) requirement has been applied as described in sub-section \ref{DR4}.}
	\label{dist}
\end{figure*}
\subsection{Mass window optimization and final statistics}
Figure \ref{bb8070} shows the $b\bar{b}$ invariant mass distributions in signal (BP2) and background events with $b$-tagging efficiencies of 80 and 70$\%$. The $t\bar{t}$ contribution is sizable in these two $b$-tagging scenarios, however, this background is suppressed well by the tight $b$-tagging scenario with average efficiency of 50$\%$. Therefore the final results are shown in Fig. \ref{dist} based on the tight $b$-tagging scenario and normalized to integrated luminosity of 1 $ab^{-1}$ which is enough for signal observation and can be easily used to re-scale the results to any other integrated luminosity. 

The final statistics can be obtained by applying a mass window cut which is optimized to achieve the maximum signal significance. Results are presented in terms of an optimized interval (window) in the distribution and the number of signal and background events in the mass window are counted for signal significance evaluation.

Table \ref{seleffDR4} shows the results of the adopted scenario presented in sub-section \ref{DR4}, including the min$\Delta R$ cut efficiency, the mass window, total signal selection efficiency, number of signal ($S$) and background ($B$) events, signal to background ratio and the signal significance (defined as $S/\sqrt{B}$), for each benchmark point, at $\sqrt{s}=1$ TeV and integrated luminosity of 1 $ab^{-1}$. The corresponding signal significance values using the first scenario (using $b_2b_3$ without min($\Delta R$) cut) are 109, 14, 31, 9.3.

The quoted results presented in Tab. \ref{seleffDR4} show that all selected benchmark points are observable, however, these results can easily be normalized to any other integrated luminosity. 

The signal cross section is independent of $\tan\beta$ as the $Z.H.A$ vertex coupling is $\sin(\beta-\alpha)$ (which was set to unity in the current analysis). The Higgs boson decays with the above assumption depend on $\cot\beta$ and below the kinematic threshold of Higgs boson decay to $t\bar{t}$, BR($H/A \rightarrow b\bar{b}$) is dominant and independent of $\tan\beta$ at tree level. Therefore, results obtained in this analysis hold for other $\tan\beta$ values which are not yet excluded by experiments.     
\begin{table*}[h!]
	\centering
	\begin{tabular}{|c|cccc|}
		\hline
		&$ BP1$ & $BP2$ & $BP3$ & $BP4$ \\
		\hline
		min$(\Delta R)$ eff.& 0.97 & 0.84 & 0.84 & 0.67\\
		\hline
		Eff. before mass win. cut & 0.25 & 0.17 & 0.15 & 0.10\\
		\hline
		Mass window [GeV] & 135-165 & 189-279 &  246-255 & 291-312 \\
		\hline
		Mass window eff. & 0.68 & 0.80 & 0.30 & 0.52\\
		\hline
		Total signal eff.  & 0.17 & 0.13 & 0.04 & 0.05\\
		\hline
		$S$ & 798 & 101 & 56 & 19\\
		\hline
		$B$ & 40 & 30 & 1.4 & 2.1\\
		\hline
		$S/B$ & 20 & 3.4 & 39 & 9.2\\
		\hline
		$S/\sqrt{B}$ & 125 & 18 & 47 & 13 \\
		\hline
		\hline
	\end{tabular}
	\caption{Selection efficiencies and the signal significance in different benchmark points based on the scenario described in sub-section \ref{DR4}. The min$(\Delta R)$ cut means that indices which minimize $\Delta R(b_ib_j)+\Delta R(b_kb_l)$ are $i=1,~j=4,~k=2~\textnormal{and}~l=3$.}
	\label{seleffDR4}
\end{table*}
\section{Conclusions}
\paragraph{}Possibility of observing CP-even and CP-odd neutral Higgs bosons, $\emph{H}$ and $\emph{A}$, was studied at a lepton collider operating at $\sqrt{s}=1$ TeV. The theoretical framework was chosen to be the type-I 2HDM with $\sin(\beta-\alpha)=1$ and $\tan\beta=10$. The Higgs boson decay to $b$-jet pair, $H/A\rightarrow b\bar{b}$, was analyzed using a fast detector simulation and two different approaches for the signal observation were presented including a kinematic correction based on the four momentum conservation. It was illustrated that a clear signal can be observed on top of the background in the $b\bar{b}$ invariant mass distribution and the signal significance exceeds $5\sigma$ in all benchmark points at integrated luminosity of $1~ ab^{-1}$. The current analysis contains improvements in several aspects, i.e., the use of beam spectrum in event generation, more dedicated ILC detector card (\texttt{ILCgen}), application of several $b$-tagging scenarios and using updated software related to theoretical and experimental considerations. Compared to the previous results reported in \cite{HtypeI_1} the observed signal distributions are much sharper and well located above the assumed Higgs boson masses, the total background is more suppressed and the signal to background ratio and the signal significance also show reasonable enhancements and more points proved to be explorable with the current analysis setup.

\end{document}